\documentclass[pre,twocolumn,superscriptaddress,floatfix,amsmath,a4paper]{revtex4-2}
\usepackage{graphicx}
\usepackage[english]{babel}
\usepackage{amsmath,amssymb}
\usepackage[utf8]{inputenc}
\usepackage{psfrag}
\usepackage{bm}
\usepackage{xcolor}
\usepackage{hyperref}
\usepackage{appendix}
\usepackage[normalem]{ulem}
\usepackage{lipsum}
\usepackage{tikz}

\begin{document}

\title{Invasion of optimal social contracts}

\author{Alessandra F. Lütz}
\email{sandiflutz@gmail.com}
\affiliation{Departamento de Física, Universidade de Minas Gerais, 31270-901, Belo Horizonte MG, Brazil}
\author{Marco A. Amaral}
\affiliation{Instituto de Humanidades, Artes e Ciências, Universidade Federal do Sul da Bahia, 45988-058, Teixeira de Freitas BA, Brazil}
\author{Ian Braga}
\affiliation{Departamento de Física, Universidade de Minas Gerais, 31270-901, Belo Horizonte MG, Brazil}
\author{Lucas Wardil}
\affiliation{Departamento de Física, Universidade de Minas Gerais, 31270-901, Belo Horizonte MG, Brazil}
\date{\today}

\begin{abstract}
The Stag-hunt game is a prototype for social contracts. Adopting a new and better social contract is usually challenging because the current one is already widely adopted and stable due to deviants' sanctions. Thus, how does a population shift from the current social contract to a better one? In other words, how can a social system leave a local social optimum configuration to achieve an optimum global state? Here, we investigate the effect of promoting diversity on the evolution of social contracts. We considered group-structured populations where individuals play the Stag-hunt game in all groups. We model the diversity incentive as a Snow-drift game played in a single focus group where the individual is more prone to adopt a deviant norm. We show that moderate diversity incentives can change the system dynamics, leading the whole population to move from the locally optimal social normal to the globally optimal one. Thus, an initial fraction of adopters of the new norm can drive the system toward the new social optimum norm. After the new social contract becomes the new equilibrium, it remains stable even without the incentive. The results are obtained using Monte Carlo simulations and analytical approximation methods.
\end{abstract}

\keywords{game theory, social norms, social dynamics}

\maketitle

\section{Introduction}

Social contracts can be defined as systems of commonly understood conventions that coordinate behavior~\cite{binmore:book:1994}. For example, social contracts contain informal rules that regulate the conduct of the citizens of a society, rules codified as Laws, moral codes, and even fashion. Societies are always envisioning better social contracts. However, it is not easy to change them due to social inertia or even because everyone is better off doing what others are already doing. Not following the rules may incur unwanted costs, and trying to change society's practices is subjected to the risk of coordination failure~\cite{STRAUB1995339,ChangingSN1}. Thus, how do better social contracts emerge? One approach to answering this question is to investigate how and why these contracts appear and survive from an evolutionary game theory perspective~\cite{skyrms_2014, Capraro2018}. 

The social contract can be modeled as a Stag-hunt game~\cite{skyrms:book:2003}. As proposed by Rousseau in \textit{A Discourse on Inequality}, ``If it was a matter of hunting a deer, everyone well realized that he must remain faithful to his post; but if a hare happened to pass within reach of one of them, we cannot doubt that he would have gone off in pursuit of it without scruple’'~\cite{russeau}. In terms of game theory, the hunting hare strategy represents the current social contract, and the hunting stag strategy represents a better social contract that has not yet been adopted. The solution where all individuals adopt the same social contract, no matter which one, is an equilibrium solution: no one has any incentive to deviate (Nash equilibrium)~\cite{Nowak2006, Szabo2007}. However, moving to a better social contract requires coordination (risky Pareto improvement). Thus, the question of how the optimal social contract emerges is translated to how society shifts from one equilibrium to the other.

Some authors have modeled the social contract as a prisoner’s dilemma game~\cite{Grujic2020,szabo_pre98,grujic_pone10,Szabo2007,Santos2006}. The state where all individuals adopt defection is interpreted as the ``state of nature'', as in Hobbes’ view of ``the war of all against all''. Cooperation is interpreted as the social contract bringing order to society~\cite{GIBBARD1974388,Hampton,kavka,kavkabook}. However, the prisoner’s dilemma does not explain how the individuals abide by the rules of the cooperative social contract since the state where everyone defects is the only equilibrium in this approach. The social contract has also been modeled as a hawk-dove game~\cite{binmore:book:1994}. The state of nature is represented by the mixed equilibrium where two strategies are played randomly. Such mixed equilibrium is akin to Hobbes’ view of the state of nature, with the strategy of randomness mimicking the chaos of the state of nature. The social contract is interpreted as one of the two pure equilibria where each individual has their role in society: one is Dove and the other Hawk. However, in this approach, the social contracts are not represented by the strategies, but by the equilibrium solutions: (hawk, dove) and (dove, hawk). In these equilibria, everyone is better off if they adopt the opposite strategy of their partner. Interestingly, behavior diversity is incorporated into the model by modeling the social contract as a Hawk-dove game, since this is an anti-polarization type of game. To sum up, although different games can be used to model social contracts, it has been argued that Stag-hunt is more appropriate~\cite{binmore:book:1994,skyrms:book:2003}.  

Social norms have been used to explain the phenomenon of human cooperation in situations where monetary-based social preferences cannot explain empirically observed behavior~\cite{capraroFoundations2021,Perc2017}. In the social preference framework, cooperative behavior is explained by assuming that the individuals seek to maximize a utility function that takes into account the payoff of everyone. However, social preference cannot explain some experiments. For example, in the Ultimatum game, the responders reject the same proposal at different rates depending on the available options to the proposer~\cite{falk2003}. This result suggests that responders follow their personal norms instead of looking exclusively at the monetary reward~\cite{Bicchieri2010}. The reader can find a comprehensive review of moral preferences in~\cite{capraroFoundations2021}. In all these works, the focus is not on the social norm evolution, but on using social norms to explain unselfishness in economic games. 

In summary, the evolution of social norms can be investigated by representing the social norm itself as the Stag-hunt game, where each strategy represents a social contract~\cite{skyrms:book:2003}. In this game, the best action is to adopt the same strategy as your partner, although one of the strategies provides a better outcome than the other. This game is an asymmetric polarization game, as everyone has the incentive to choose the strategy adopted by the majority of the population, even if this strategy is not the global optimum.

The payoff structure of the Stag-hunt game does not answer how a society can jump from the local equilibrium with a low payoff to a better one. Additional mechanisms must work to allow the jump. For example, the optimal equilibrium is facilitated if the agents exchange costly signals~\cite{skyrms2002,SANTOS2011}. More specifically, if the game is a Stag-hunt game with strategies $A$ and $B$, and the agents can send signals ($1$ or $2$),  the individuals can take action according to a rule (if the signal is $1$ they adopt $A$, otherwise they adopt $B$). Notice that although no individual has previous information about the meaning of the sign, the communication introduces some correlations that facilitate the emergence of the better but riskier equilibrium~\cite{skyrms2002}. The social structure can also modify the nature of the game. If the agents control their interactions, individuals with the same strategy can interact preferentially and promote the new norm~\cite{watts2001}.

Adopting a new behavior that does not conform to the currently adopted norm can be quite challenging. Individuals are part of social groups that exerts pressure on them~\cite{young2015evolution}. Nevertheless, there may be groups where the individuals feel more comfortable displaying new behaviors and conforming to whatever they believe. Promoting a culture of diversity is shown to be an escape from the trap of a bad equilibrium~\cite{Szolnoki2018a,cheng2011promoting,cheng2014population,squillero2016divergence,qin2018diversity,Sendina-Nadal2020,Amaral2020, Santos2008,santos_jtb12,Amaral2018b}. 
Incentives promoting diversity can be mapped to the Snow-drift payoff structure, where adopting a different strategy from your partner is the best choice. Here we adopt the metaphor of the Snow-drift game instead of Hawk and Dove to avoid any allusion to the antagonistic interaction present in the Hawk and Dove metaphor. Suppose the decision problem of which social contract one should adopt is determined by the payoff structure of a combination of the social contract and diversity promotion policy. In that case, we may ask to what extent the promotion of diversity (Snow-drift game influence) can shift the whole population toward the optimal social contract.

Here, we investigate how incentives promoting diversity affect the dynamics of social contracts. We model the social contract as a Stag-hunt game and the diversity incentive as a Snow-drift game. The population is structured in groups. Each individual pertains to a fraction of them, where the Stag-hunt game is played. There is one group where the individual is encouraged to adopt a strategy that does not conform to the majority. This encouragement is modeled as a Snow-drift game. The strategies evolve according to imitation dynamics, where the strategies that yield higher payoffs spread at higher rates. We show that the equilibrium is shifted to the optimum social norm for moderate Snow-drift incentives. Because it is a new equilibrium, the new social contract is stable even without the diversity incentive. However, if the individuals interact with many others in different groups, the overall social pressure can demand more substantial Snow-drift incentives. We also show that the same results are observed in populations structured in square lattices, where the Stag-hunt and the Snow-drift games are played within a range of neighbors. Our analysis is supported by analytical approximations that allow us to explain the results using simple concepts of game theory. 

\section{Model}

The population is structured in groups, and the individuals can have multiple group affiliations. More specifically, there are $n$ groups in a population of size $N$. Each individual pertains to one group, which we call the focus group, and to a fraction $q$ of the other non-focal groups. The focus groups are defined by initially setting $N/n$ individuals to each group. 

The social norms are modeled as strategies in a Stag-hunt game, which is played in all groups. The norm that yields the social optima is represented by $A$ and the other one by $B$. Interactions of type $A-A$ yield a payoff equal to $1$, while $B-B$ yields $0$. If individuals with different norms interact, $A$ receives  $-\delta$ ($0<\delta<1$) and $B$ receives $1-r$ ($0<r<1$). The game is specified by the parameter set $(r, \delta)$ and is represented by the following payoff matrix:
\begin{align}
\mbox{Stag-hunt}=\begin{pmatrix}
1 & -\delta \\
1-r & 0\\
\end{pmatrix}
\end{align}
The best solution is to adopt the same strategy as the partner. Because $r>0$, the norm $A$ is the social optimum. However, if everyone is adopting $B$ and a single individual moves to the social optimum $A$, this individual faces the risk of receiving $-\delta$, which is worse than $0$. 

The diversity incentive is modeled as a Snow-drift game, which is played only in the focus group of each individual. The payoffs from Snow-drift are parameterized by $(r', \delta')$, with $0<r'<1$ and $0<\delta'<1$, and is represented by the payoff matrix:
\begin{align}
\mbox{Snow-drift}=\begin{pmatrix}
1 & \delta' \\
1+r' & 0 \\
\end{pmatrix}
\end{align}
The best solution is to adopt the opposite strategy of the partner. Notice that in a $(A, B)$ interaction, the individual adopting $B$ receives $1+r'$, which is larger than the payoff that $A$ receives, which is $\delta'$. That is, the diversity incentive is not benefiting the social contract $A$. As a last remark, individuals in $(A, A)$ interactions receive more than those in $(B, B)$, but this is not an issue because the diversity incentive is most important when strategy $A$ is the minority.

The cumulative payoff that player $f$ obtains in each game is: 
\begin{eqnarray}
\pi_{SH,f}&=&\sum_{(f,z)}\pi_{f,z} \\
\pi_{SD,f}&=&\alpha\sum_{\left<f,z\right>}\pi_{f,z}
\end{eqnarray}
where $(f, z)$ indicates that $f$ and $z$ are connected and $<f, z>$ indicates that $f$ and $z$ belong to the same focus group.
The weight of the Snow-drift is controlled by the parameter $\alpha$ such that the total payoff of player $f$ is given by 
\begin{equation}
\pi_f=\pi_{SH,f}+\pi_{SD,f}.
\end{equation}
Notice that the higher the number of groups connected to an individual, the higher the weight of Stag-hunt on the total payoff. 

\begin{figure*}[ht]
\begin{center}
\includegraphics[width=0.7\linewidth]{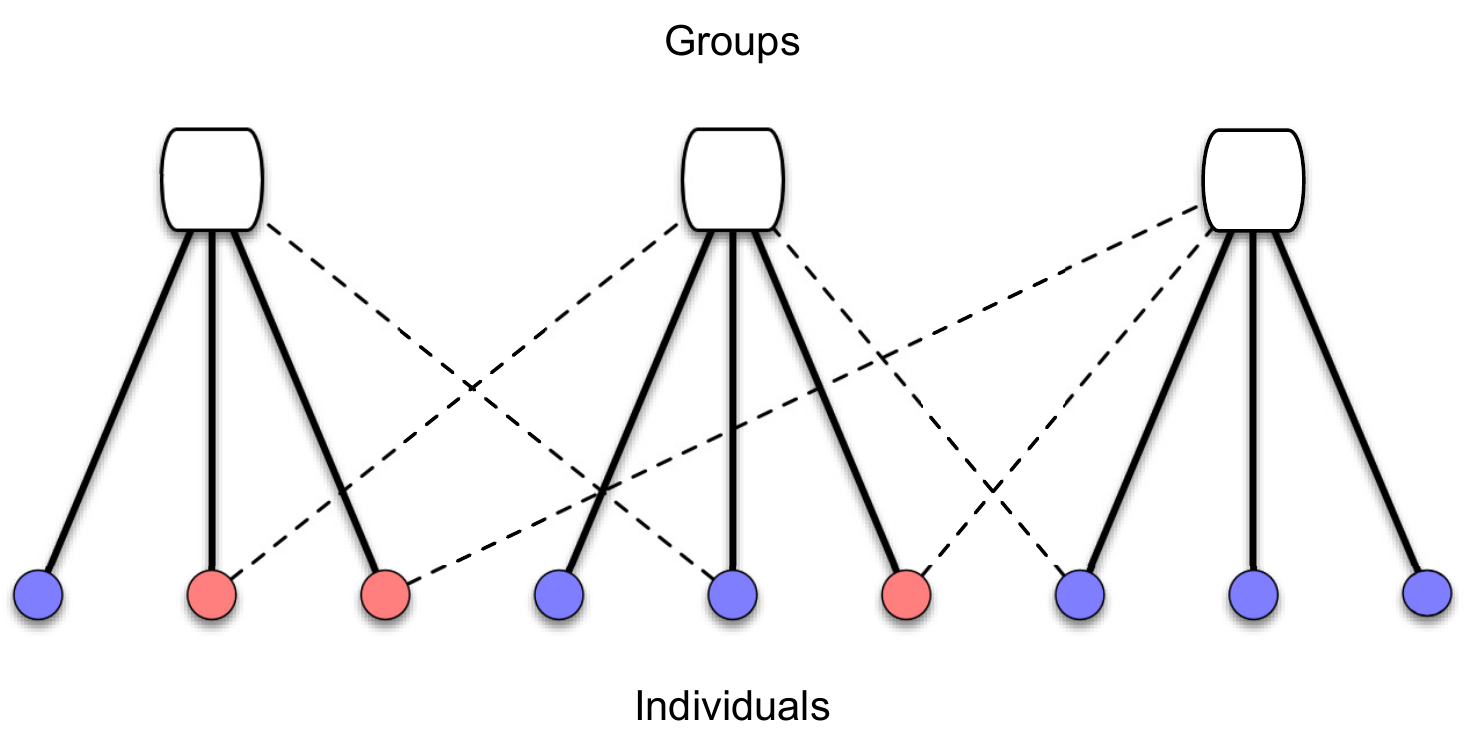}
\caption{Group structure. The individuals are represented by circles and the groups by rounded squares. Each individual belongs to one focal group (continuous line) and is connected to each one of the other groups with probability $q$ (dashed line). The two social norms, $A$ and $B$, are represented by the colors, blue and red, respectively. Here, $N=9$, $n=3$, and $q=1/3$.}
\label{fig.groupstructure}
\end{center}
\end{figure*}

The evolution of strategies is determined by imitation dynamics. A random player $f$ is selected to compare his payoff with a random player $z$, connected to him. If they have different strategies, the first player imitates the strategy of the second one with a probability given by the Fermi rule:
\begin{equation}
    P_{f\leftarrow z}=\frac{1}{1+\exp\left(-(\pi_z-\pi_f)/K\right)},
\end{equation}
where $K$ is the selection intensity representing the population level of irrationality. Each generation consists of $N$ repetitions of the imitation step.

\section{Results}

The imitation dynamics can be analyzed using a mean-field approximation. For the simple case of one group ($n=1$), the mean-field equation becomes
\begin{equation} \label{n0meanfield}
    \frac{dx}{dt}=(1-x)x\sinh{\left(\pi_A(x)-\pi_B(x)\right)},
\end{equation}
where $x$ is the fraction of individuals adopting $A$, while $\pi_A$ and $\pi_B$ are the average payoffs of individuals adopting $A$ and $B$, respectively (see the Appendix~\ref{Appendix}). 
For $n=1$ there is a single well-mixed group where all individuals play the two games. Thus, the cumulative payoffs are effectively determined by the sum of the matrices of the two games:
\begin{equation} \label{summatrix}
M=\begin{pmatrix}
1+\alpha & -\delta+\alpha \delta' \\
1-r+\alpha (1+r') & 0
\end{pmatrix}
\end{equation}

Let us first recall the main results for the Stag-hunt dynamics by setting $\alpha=0$. The analysis of Eq.~\ref{n0meanfield} shows that
there is an unstable equilibrium $x^*$ (solution of ${\pi_A(x)-\pi_B(x)=0}$). Therefore, if the fraction of $A$ at time $t$ is such that $x<x^*$, because $\pi_A(x)-\pi_B(x)<0$ for $x$ in that range, the population goes to the state $x=0$ (all $B$). If $x>x^*$, the population goes to the state $x=1$ (all $A$). Thus, if a small amount of $A$ (a fraction $\epsilon$)  invades a population initially at $x=0$, as long as ${\epsilon <x^*}$, the invader has no chance. The invasion must be large enough to overcome the invasion barrier determined by the unstable equilibrium $x^*$. 

On the other side, in the dynamics of a Snow-drift game new strategies can always invade. The analysis of Eq.~\ref{n0meanfield} for very large $\alpha$, when the game is a pure Snow-drift, shows that there is a stable equilibrium state, where $A$ and $B$ coexist. Thus, a small amount of $A$ invaders have the incentive to maintain their strategies. 

In the combination of both the Stag-hunt and Snow-drift, the effect of increasing the incentive provided by the Snow-drift game, which is parameterized by $\alpha$, is to change the dynamics from that of the Stag-hunt to that of the Snow-drift. If $\alpha$ is low, the weight of the Stag-hunt payoff is larger, and the optimum strategy $A$ cannot invade. If $\alpha$ is too large, the dynamics change completely, being determined by the Snow-drift payoff, where both strategies coexist. However, if $\alpha$ is moderate, more specifically, if
\begin{equation} 
    \frac{\delta}{\delta'}<\alpha < \frac{r}{r'},\label{cond-singlegame}
\end{equation}
then, not only $A$ can invade, but it will certainly dominate the population. Equation~\ref{cond-singlegame} is obtained by a simple Nash equilibrium analysis of the payoff matrix in Eq.~\ref{summatrix}. For $\alpha$ in this interval, the payoff structure becomes equivalent to that of a Harmony game, where $A$ is a global attractor of the dynamics. If the goal of the Snow-drift incentive is to shift the population to the social optimum without ending at a coexistence equilibrium, then the moderate $\alpha$ solution is the best. Thus, if the incentive provided by the Snow-drift is moderate, the optimal social contract can invade the population and persists even if the Snow-drift incentive is turned off.  

\begin{figure*}[ht]
\begin{center}
\begin{tikzpicture}
    \node[anchor=south west,inner sep=0] (image) at (0,0) {\includegraphics[width=0.5\textwidth]{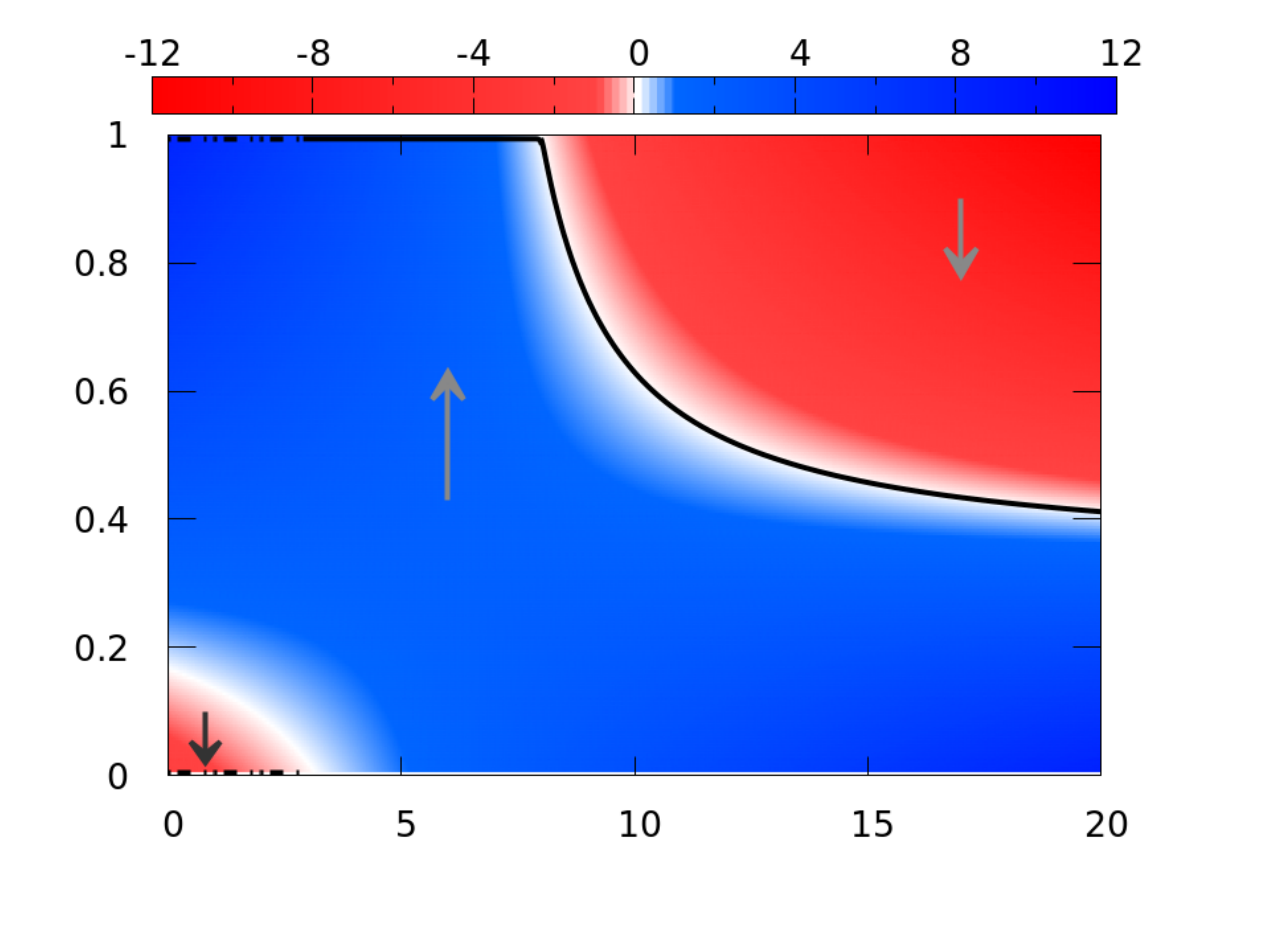}\includegraphics[width=0.4\textwidth]{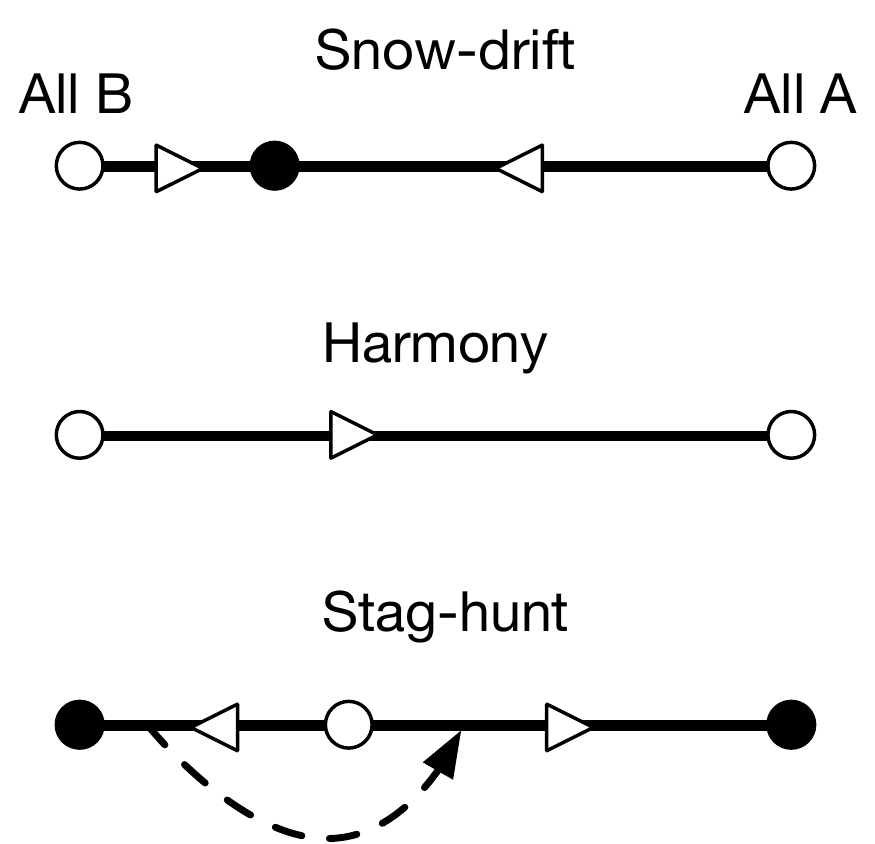}};
    \begin{scope}[
        x={(image.south east)},
        y={(image.north west)}
    ]
            \node [font=\rmfamily, rotate=90] at (0.,0.5) {\large X};
            \node  at (0.26,0.08) {\Large $\alpha$};
            \node  at (0.28,1) {\Large $\pi_A-\pi_B$};
    \end{scope}
\end{tikzpicture}
\caption{Replicator dynamics of the average game. The first diagram, on the left, shows the payoff difference $\pi_A-\pi_B$ in the replicator equation, as a function of the fraction of $A$ and the Snow-Drift incentive $\alpha$. The arrows indicate the sign of $\dot{x}$ (positive, if the arrow points up, and negative otherwise). The second diagram, on the right, shows the three main regions of the first diagram using a simplex representation for the strategies' fractions. For low values of $\alpha$, Stag-hunt dominates and there is an unstable equilibrium state around $x^*\simeq0.1$. All agents end up adopting strategy $A$ or $B$, depending on whether $x$ is lower or higher than $x^*$. The dashed arrow represents the invasion barrier that $A$ would have to overcome if most of the agents adopt strategy $B$. For intermediary $\alpha$, the system behaves as a Harmony game, and everyone adopts strategy $A$. Finally, for high values of $\alpha$, Snow-drift dominates, and strategies $A$ and $B$ coexist. Here, $r=0.5$, $\delta=0.1$, $r'=1$ and, $\delta'=0.5$.}
\label{fig.payoffdiff}
\end{center}
\end{figure*}

Still in the $n=1$ case, the unstable equilibrium of Stag-hunt determines the invasion barrier for the norm $A$: the higher the value of $x^*$, the harder it is for $A$ to invade. Figure~\ref{fig.payoffdiff} shows how the payoff difference $\Pi_A(x)-\Pi_B(x)$ changes as the fraction $x$ of $A$ and the incentive $\alpha$ vary. 
If there is no Snow-drift incentive ($\alpha=0$), only a massive conversion of $A$ will drive the population to the norm $B$. However, if a moderate incentive is provided, any initial fraction of $A$ will convert the population. As expected, if $\alpha$ is too large, the dynamics changes and coexistence will be the final state independently of the initial conditions.

In the general case where the population is split into groups, a close look at the general mean field approximation (discussed in the Appendix~\ref{Appendix}) shows that the fraction of $A$ in each group tends to be close to each other. Thus, because all groups have roughly the same fraction of $A$ at any time, the analysis can be simplified to the analysis of a single group, the  $n=1$ case, with the average payoff matrix given by
\begin{equation}
M=\begin{pmatrix}
\tilde{q}+\alpha & \tilde{q}S+\alpha S' \\
\tilde{q}T+\alpha T' & 0
\end{pmatrix},
\end{equation}
where  $\tilde{q}=1+q(n-1)$ is the number of groups where the Stag-hunt is played. The dynamics are now determined by the impact of the Stag-hunt payoff relative to the Snow-drift payoff, which is controlled by the parameters $\alpha$ and $q$. 

First, if the group structure does not change, that is, if $q$ is fixed, we would like to find moderate values of the incentive $\alpha$ that turn the Stag-hunt payoff into a Harmony game payoff. The condition is given by 
\begin{equation}
    \tilde{q}\frac{\delta}{\delta'}<\alpha < \tilde{q}\frac{r}{r'}.
\end{equation}
On the other side, if $\alpha$ is fixed, then the effect of the Snow-drift incentive depends on the group structure. If the individuals play Stag-hunt in many groups (high $q$), then the social pressure overcomes the Snow-drift incentive and it is nearly impossible for $A$ to invade. More specifically, let us fix $\alpha$, with $\alpha>r/r'$, so that if $\tilde{q}=1$ the Snow-drift is dominant. In this case, increasing $\tilde{q}$ means that more Stag-hunt games are played. Thus, if $\tilde{q}$ is too large the Stag-hunt dynamics is dominant. However, if
\begin{equation}
\alpha \frac{r'}{r}<\tilde{q}<\alpha \frac{\delta'}{\delta},
\end{equation}
then the effective game is the Harmony game and the strategy $A$ can invade and dominate. In other words, only if the number of groups is moderate the new social contract can invade.

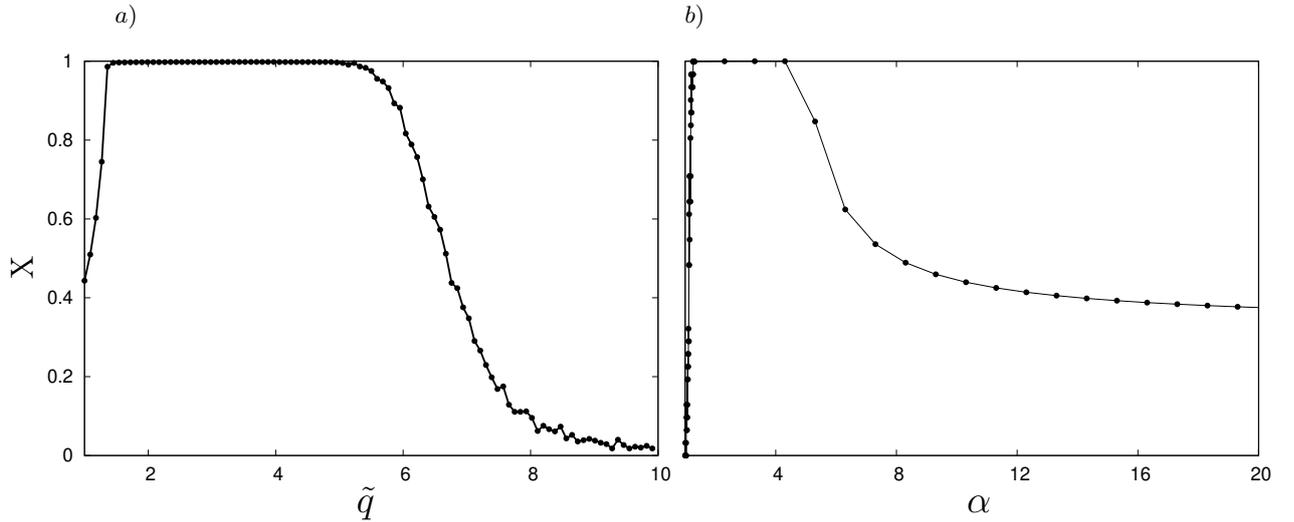
\begin{figure*}[ht]
\begin{center}
\hspace{-6cm}$a)$\hspace{7.2cm}$b)$

\vspace{3mm}

\begin{tikzpicture}
        \node[anchor=south west,inner sep=0] (image) at (0,0) {\includegraphics[width=0.9\textwidth]{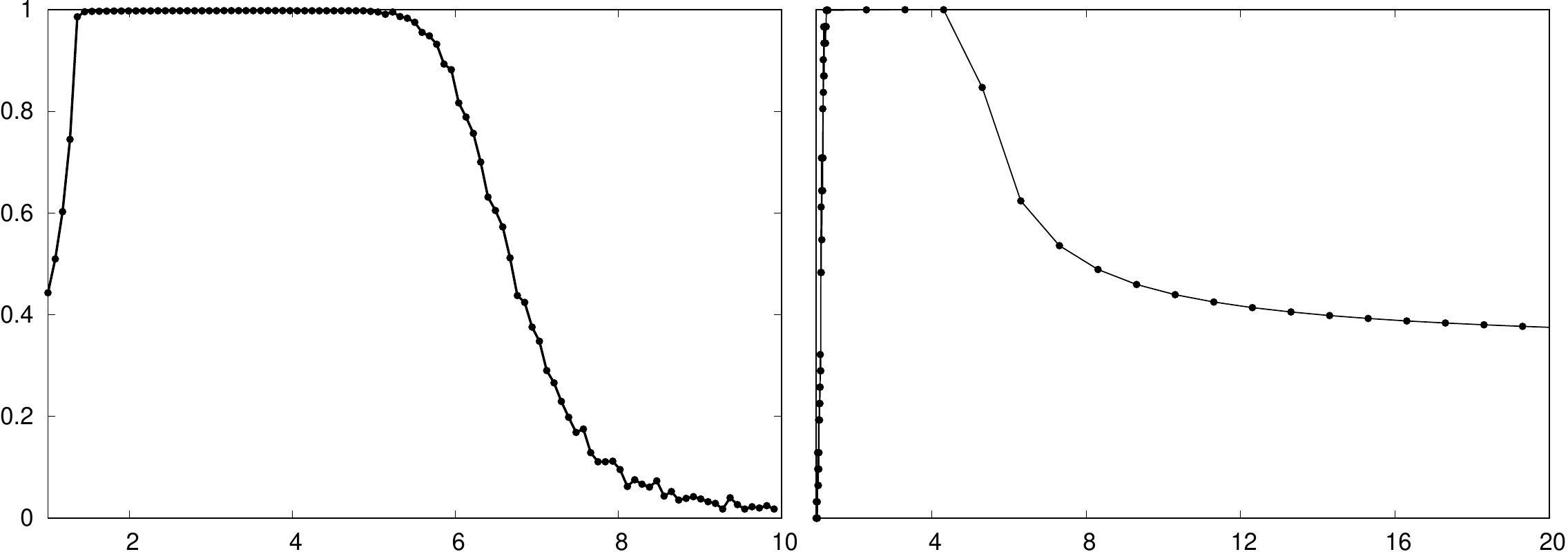}};
    \begin{scope}[
        x={(image.south east)},
        y={(image.north west)}
        ]
            \node [font=\rmfamily, rotate=90] at (-0.02,0.5) {\large X};
            \node  at (0.76,-0.05) {\Large $\alpha$};
            \node  at (0.26,-0.05) {\Large $\tilde{q}$};
    \end{scope}
\end{tikzpicture}
\caption{Average fraction of the number of individuals adopting the optimal social contract, $x$. Figure $a$ shows the fraction $x$ as a function of the number of groups connected to each agent, $\tilde{q}=1+(n-1)q$, while figure $b$ shows that fraction as a function of the incentive $\alpha$. Here, $r=0.5$, $\delta=0.1$, $r'=1$ and, $\delta'=0.5$. The initial fraction of $A$ is $x(0)=0.1$, the number of agents is $N=10000$ and the number of groups is $n=10$. In figure $a$, the Snow-drift incentive is $\alpha=1$.}
\label{fig.dXq_dXa}
\end{center}
\end{figure*}

The simulations corroborate the mean-field analysis, as shown in Figs.~\ref{fig.dXq_dXa}$a$ and~\ref{fig.dXq_dXa}$b$. Figure~\ref{fig.dXq_dXa}$a$ shows the equilibrium fraction of $A$ (the globally optimal social norm) as a function of the number of groups connected to each agent, $\tilde{q}=1+(n-1)q$, while in Fig.~\ref{fig.dXq_dXa}$b$ the fraction of $A$ is shown as a function of the Snow-drift incentive $\alpha$. The optimum influence of the incentive $\alpha$ over the success of $A$ is felt for a moderate number of groups since $x$ is at its maximum value for intermediate $\tilde{q}$ (see Fig.~\ref{fig.dXq_dXa}$a$). If the system is highly connected, Stag-hunt dominates, and the social norm $A$ disappears when $x(0)$ is low. Notice that the incentive $\alpha$ has to be low to keep Snow-drift from dominating the dynamics, as shown in Fig.~\ref{fig.dXq_dXa}$b$.

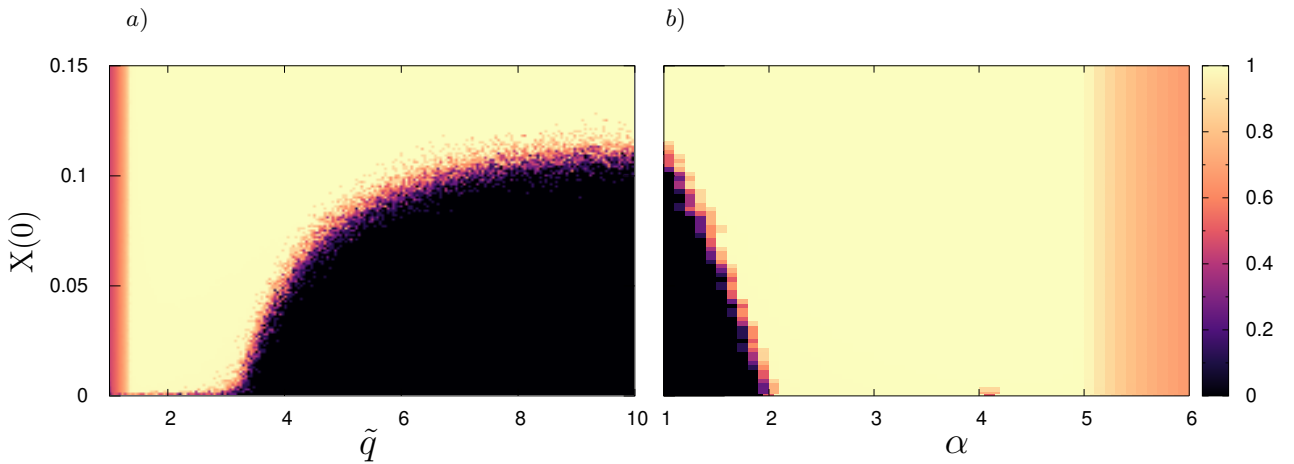
\begin{figure*}[ht]
\begin{center}

\hspace{-6.1cm}$a)$\hspace{6.8cm}$b)$

\vspace{3mm}

\begin{tikzpicture}
        \node[anchor=south west,inner sep=0] (image) at (0,0) {\includegraphics[width=0.9\textwidth]{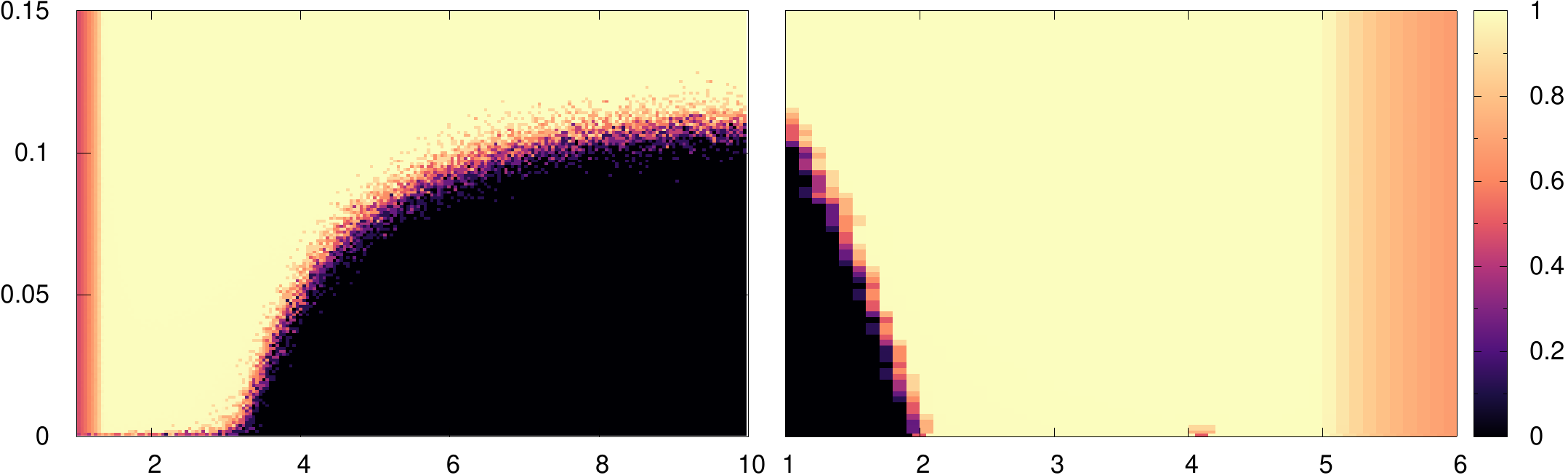}};
    \begin{scope}[
        x={(image.south east)},
        y={(image.north west)}
        ]
            \node [font=\rmfamily, rotate=90] at (-0.02,0.5) {\large X(0)};
            \node  at (0.74,-0.05) {\Large $\alpha$};
            \node  at (0.26,-0.05) {\Large $\tilde{q}$};
    \end{scope}
\end{tikzpicture}
\caption{Diagram for the fraction $x$ of individuals adopting the optimal social contract. Figure $a$ shows $x$ as a function of its initial value $x(0)$, and the number of groups connected to an individual $1+(n-1)q$, while in figure $b$ the fraction $x$ is shown as a function of $x(0)$ and the Snow-drift incentive $\alpha$. Here, $r=0.5$, $\delta=0.1$, $r'=1$ and, $\delta'=0.5$.}
\label{fig.diagQxA}
\end{center}
\end{figure*}

The effect of the initial fraction of $A$, $x(0)$, is shown in Figs.~\ref{fig.diagQxA}$a$ and~\ref{fig.diagQxA}$b$. One can see that the initial condition $x(0)$ affects the equilibrium fraction of $A$ when the dynamics are dominated by Stag-hunt, which happens if the number of groups, $\tilde{q}$, is high enough (Fig.~\ref{fig.diagQxA}$a$), and if the Snow-drift coefficient $\alpha$ is sufficiently low (Fig.~\ref{fig.diagQxA}$b$). Additionally, the minimum value of $x(0)$ necessary for preventing $A$ from being extinguished increases with the number of connections $\tilde{q}$, especially when the system goes from being poorly connected ($\tilde{q}\leq3$) to being moderately connected ($3<\tilde{q}<5$). This effect comes from the majority's strategy strongly influencing agents' strategies in Stag-hunt. Thus, when an agent plays Stag-hunt with most of its connections, he is more susceptible to adopting the majority's strategy. However, the influence of $x(0)$ for the persistence of $A$ weakens as the Snow-drift incentive $\alpha$ increases, i.e., when the local interactions that encourage different opinions become more important, as can be seen in Fig.~\ref{fig.diagQxA}$b$.  

\begin{figure*}[ht]
\begin{center}

\begin{tikzpicture}
        \node[anchor=south west,inner sep=0] (image) at (0,0) {\includegraphics[width=0.9\textwidth]{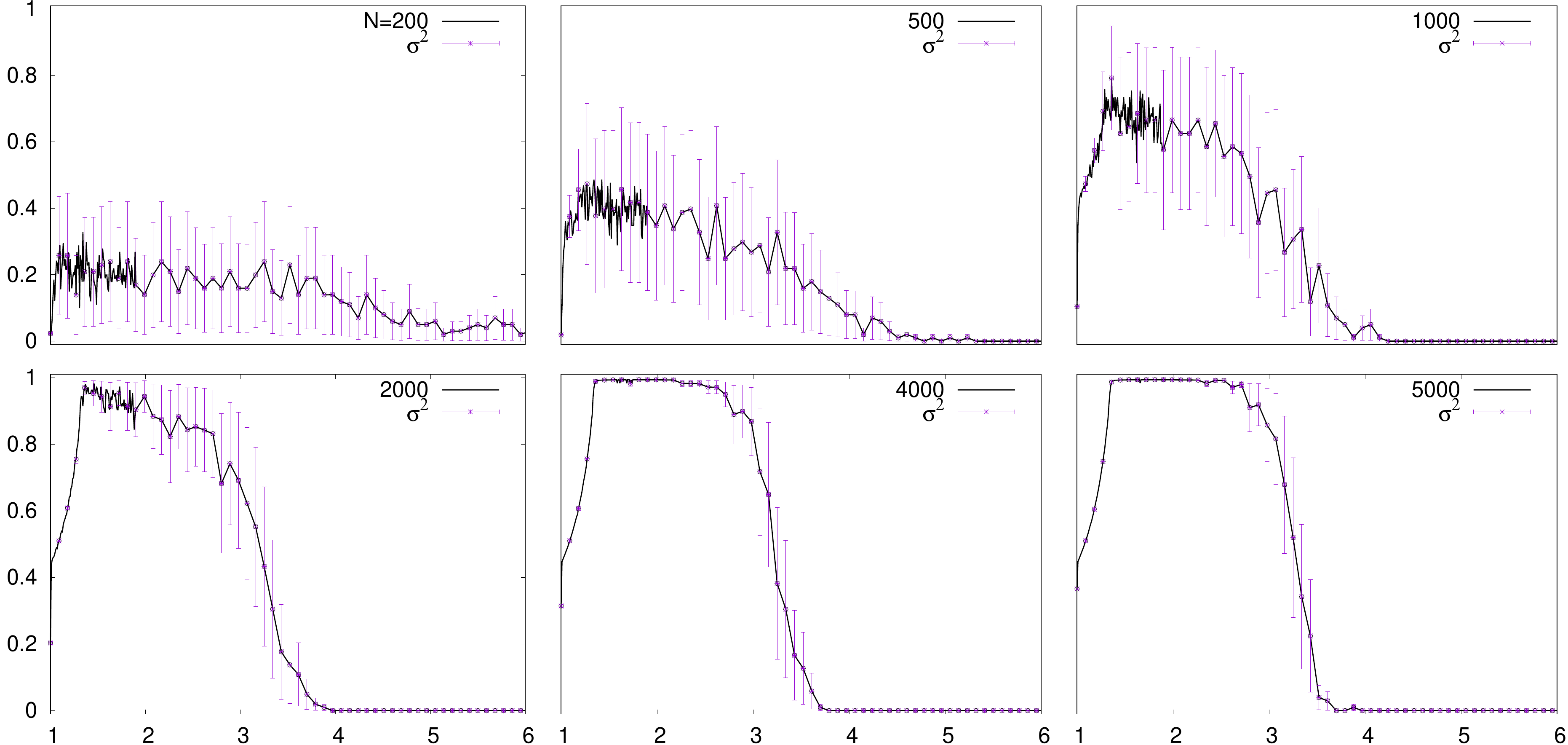}};
    \begin{scope}[
        x={(image.south east)},
        y={(image.north west)}
        ]
            \node [font=\rmfamily, rotate=90] at (-0.02,0.5) {\large X};
            \node  at (0.5,-0.05) {\Large $\tilde{q}$};
    \end{scope}
\end{tikzpicture}

\caption{Average fraction $x$ of individuals adopting the optimal social contract for different system sizes. Each figure shows $x$ as a function of the number of connections per agent, $1+(n-1)q$, for a different system size. Here, the number of groups is $n=10$, the initial fraction of $A$ is $x(0)=0.01$ and the Stag-hunt and Snow-drift parameters are $r=0.5$, $\delta=0.1$, $r'=1$ and $\delta'=0.5$. Notice that despite the finite size effects, strategy $A$ is still facilitated for moderate values of $\tilde{q}$.}
\label{fig.multi_dXq}
\end{center}
\end{figure*}

The mean-field approximation is valid only if the size of the groups is large. If small, fluctuations play a major role and the mean-field approach is not precise. Even so, we see that strategy $A$ is facilitated for moderate values of $q$, as shown in Fig.~\ref{fig.multi_dXq}, for all population sizes. Notice that for small population sizes, the average values in Fig.~\ref{fig.multi_dXq} represent fixation probabilities and not stationary values. The reason is that for small sizes, the states where all individuals are $A$ or all are $B$ are reached with probability one independently of the game being played. The coexistence stationary solution is meta-stable even if only the Snow-drift game is played. The mean-field behavior is recovered for the large groups.

To further investigate the robustness of our results, we also consider a square lattice version of our model. Each agent occupies a site in a square lattice and plays Stag-hunt and Snow-drift with all the sites within a range of $R_{SH}$, for Stag-hunt, and $R_{SD}$, for Snow-drift, with $R_{SH}\ge R_{SD}$. The distance between two neighboring sites is set to be $1$. The interaction ranges delimit groups and have a similar role as the parameter $q$ in the previous version of the model. 

\begin{figure*}[ht]
\begin{center}

\hspace{-6cm}$a)$\hspace{7.2cm}$b)$

\vspace{3mm}

\begin{tikzpicture}
        \node[anchor=south west,inner sep=0] (image) at (0,0) {\includegraphics[width=0.9\textwidth]{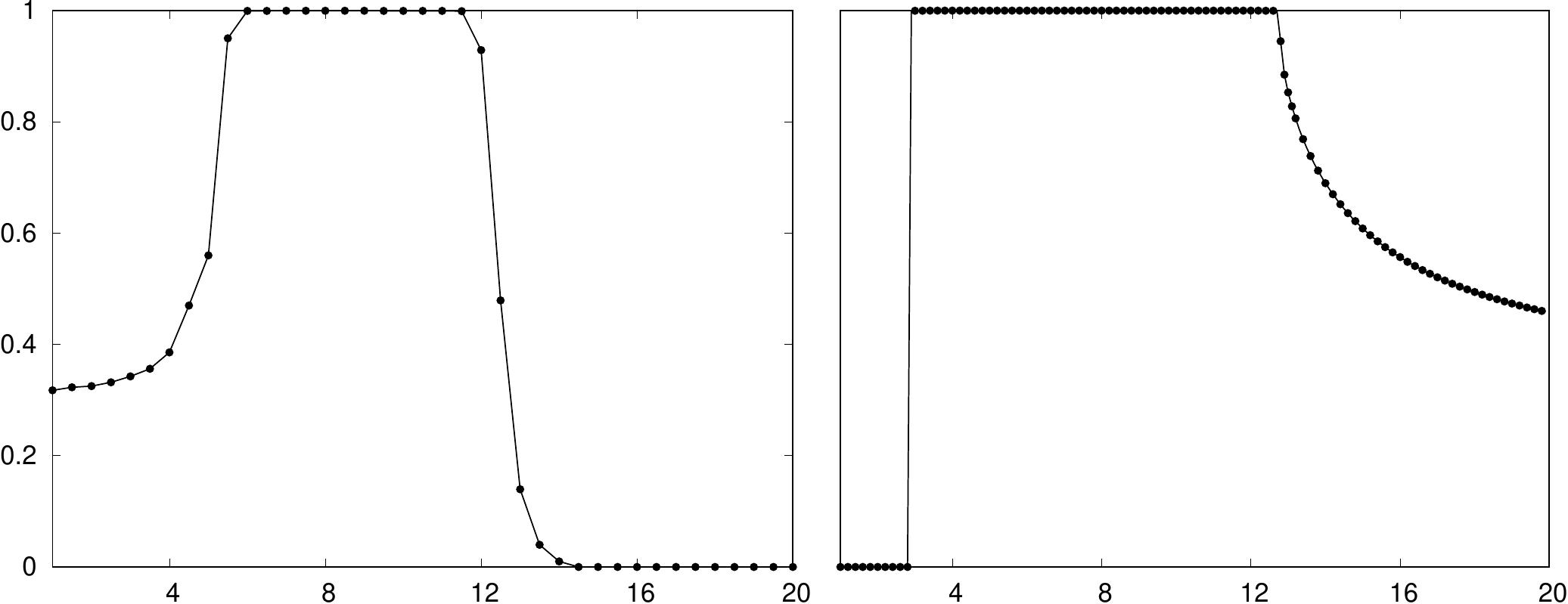}};
    \begin{scope}[
        x={(image.south east)},
        y={(image.north west)}
        ]
            \node [font=\rmfamily, rotate=90] at (-0.02,0.5) {\large X};
            \node  at (0.26,-0.05) {\Large $R_{SH}$};
            \node  at (0.74,-0.05) {\Large $\alpha$};
    \end{scope}
\end{tikzpicture}

\caption{Average fraction $x$ of individuals adopting the optimal social norm $A$ in the square lattice model. Figure $a$ shows $x$ as a function of the Stag-hunt influence radius, $R_{SH}$, while in figure $b$ the fraction $x$ is shown as a function of the Snow-drift incentive $\alpha$. Here, the influence radius for the Snow-drift game in both figures is $R_{SD}=4$ and the initial fraction of $A$ is $x(0)=0.1$. The Stag-hunt e Snow-drift parameters are  $r=0.5$, $\delta=0.1$, $r'=1$ and, $\delta'=0.5$. In figure $b$, the influence radius for Stag-hunt is $R_{SH}=20$.}
\label{fig.rede_dXr_dXa}
\end{center}
\end{figure*}

In the square lattice version, the agents have weaker connections, since the clustering coefficient is always lower for this network than in the group-structured population. This difference could impact the results since the network clustering affects the spreading ability of the social norms. Despite such differences, the results for both settings are very similar.
Figures~\ref{fig.rede_dXr_dXa}$a$ and~\ref{fig.rede_dXr_dXa}$b$ show, respectively, the fraction of $A$ as a function of the Stag-hunt influence radius $R_{SH}$ and the fraction of $A$ as a function of the Snow-drift incentive $\alpha$.
Similar to the results for the group simulation (see Fig.~\ref{fig.dXq_dXa}), the incentive $\alpha$ most benefits the norm $A$ when Stag-hunt influence is limited but relevant. If $R_{SH}$ is way larger than the Snow-drift influence radius $R_{SD}$, then Stag-hunt dominates, and $A$ disappears for low $x_A(0)$. 
Additionally, the incentive $\alpha$ most benefits $A$ when it is not too large, and the system is not dominated by Snow-drift.

\section{Conclusions}
Once established, social contracts tend to stay because of the very nature of the incentives: deviation from the norm is not beneficial. Thus, how new equilibrium can emerge? Even if the new equilibrium is better, overcoming the invasion barrier is a problem. Here we show that the new equilibrium can be reached by providing a low incentive for diversity. 

Our model assumes that agents can maintain a focus group (akin to close friends and trustworthy contacts) where the individuals are more prone to deviate and adopt a new norm. The new norm can be stimulated through incentives for diversity and inclusion. By modeling the social contract as a Stag-hunt and the incentives as a Snow-drift game, we show that the optimum social norm can invade if the incentive provided by the Snow-drift game is just moderate.

The Snow-drift incentive can be temporary. After the population is driven to the new optimum equilibrium, where all adopt the new norm, the incentive can be turned off because the stability of the social norm will play in favor of the new norm. If the Snow-drift incentive is too large, the new norm will coexist with the old one, and it is not guaranteed that the population will move toward the socially optimum norm after turning off the incentive.

The Stag-hunt payoff structure assumes that there is a social optimum that is the best outcome for everyone. We do not analyze here group-specific social contracts, so different groups are better off by adopting different social norms. Moreover, respect for diversity is coded as a moral rule in many societies. By turning off the Snow-drift incentive after the new social optimum is reached, the model is not against such moral rules. After all, we assume that it is in the best interest of everyone to move to the new equilibrium. Instead, we see that diversity policies can help to coordinate actions.

Changing social norms is undoubtedly a complex matter. We do not claim that there is always a simple incentive structure that can shift the equilibrium to the optimum social norm. The point of our work is to shed light on the backbone of incentive structure: social norms do not favor deviant behavior, and policies for diversity and inclusion facilitate the coexistence of different norms. Using an example from physics, if a block is at rest on an inclined plane sustained by the static frictional force, we know that, although the block is not moving, forces are acting on the block. Thus, even if it is very likely that many factors contribute to the evolution of the social contract, we show that incentives for diversity and inclusion, if moderate, may drive the population toward the social optimum.

\appendix

\section{Appendix}
\label{Appendix}

The justification for assuming that the fraction of $A$ in all groups is approximately the same is derived as follows. Let $N_{Ai}$ be the number of individuals of norm $A$ that belong to group $i$. If $n$ is the number of groups, each group has $N/n$ individuals, and $N_{Ai}$ ranges from $0$ to $N/n$. The following analysis uses the fact that the payoff matrix in the focal group is the sum of the payoff of the two games. Let $(a,b,c,d)=(1,-\delta,1-r,0)$ be the Stag-hunt payoff matrix and $(a',b',c',d')=(1+\alpha,-\delta+\alpha \delta',1-r+\alpha(1+r'),0)$ the sum of the payoff of the two games.  

In our approach, we consider that each individual in the focus group $i$ has a probability $q$ of being connected to each of the remaining $n-1$ groups. Let $x_i = N_{Ai}/N$, with $0\le x_i\le 1/n$, be the fraction of $A_i$ individuals in the total population. Because the groups have well-mixed interactions, the payoffs gained by $A$ and $B$ individuals belonging to the same focus group $i$ are given by
\begin{eqnarray*}
\pi_{Ai}&=& a'nx_i + b'(1-nx_i) + q\sum_{j \neq i} anx_i + b(1-nx_i) \\
\pi_{Bi} &=& c'nx_i + d'(1-nx_i) + q\sum_{j \neq i} cnx_i + d(1-nx_i).
\end{eqnarray*}

The next step is to approximate the transition rates. Recall that all individuals have an equal probability of being chosen to change their strategy. Thus, an $A_i$ individual has a probability $x_i$ of being chosen, and a $B_i$ individual has a probability $(1/n - x_i)$. Also, since each individual is certainly linked with the members of the same focus group, the probability that he compares his payoff with one of the $Ai$ is proportional to $x_i$,  and with one of the $Bi$ is proportional to $(1/n - x_i)$. For the non-focal groups, the probability that an individual from $i$ compares its strategy with an $Aj$, or a $Bj$, individual is proportional to $qx_{j}$,  or $q(1/n - x_j)$, respectively. Therefore, the transition rates can be written as
\begin{eqnarray*}
T^{+}_{i}&=& \frac{1}{Z}\left( \frac{1}{n} - x_{i}\right) \sum_{j} q_{ij} x_{j} \frac{1}{1+e^{\pi_{Bi} - \pi_{Aj}}} \\
T^{-}_{i} &=& \frac{1}{Z}x_{i} \sum_{j} q_{ij} \left( \frac{1}{n} - x_{j}\right) \frac{1}{1+e^{\pi_{Ai} - \pi_{Bj}}},
\end{eqnarray*}
where $q_{ii} = 1$, $q_{ij} = q$ for $i\neq j$, and $Z$ is a normalization factor. For large populations, the system is mainly driven by the drift term, and it can be described by the following deterministic set of rate equations:
\begin{eqnarray*}
\frac{dx_i}{d\tau} &=& \sum_{j} q_{ij}\left[ \frac{\frac{1}{n} (x_{j} - x_{i}) + \left( \frac{1}{n} - x_{i}\right)x_{j}e^{\pi_{Ai} - \pi_{Bj}}}{(1+e^{\pi_{Ai} - \pi_{Bj}})(1+e^{\pi_{Bi}  \pi_{Aj}})}\right.\\
&-&\left.\frac{\left( \frac{1}{n} - x_{j}\right)x_{i}e^{\pi_{Bi} - \pi_{Aj}}}{(1+e^{\pi_{Ai} - \pi_{Bj}})(1+e^{\pi_{Bi} - \pi_{Aj}})}\right], 
\end{eqnarray*}
where we have rescaled the time. Notice that for $n=1$ Eq.~\ref{n0meanfield} in the main text is recovered.

Finally, we see that if $\pi_{Ai} \approx \pi_{Bj}$ for all $i,j$, which is the case in the regime of weak selection, then $e^{\pi_{Bi} - \pi_{Aj}} \approx 1 $ for all $i,j$, and the set of rate equations drives the variables $x_{i}$ near to each other in the first order. The reason is that if $x_i<x_j$ for some $j$, then the first order term $(x_{j} - x_{i}) $ contributes with a positive term to the equation, with the opposite happening if $x_i>x_j$. Thus, the overall effect is that the variables $x_i$ will evolve in time close to each other for all $i$. This is interesting because we can describe the system only in terms of the total fraction of $A$ types, $x = \sum_{i}x_i$, since in this approximation we have $x_{i}\approx x/n$ after a short relaxation time.

\bibliographystyle{ieeetr}

\end{document}